\documentclass[12pt]{article} % modifi\'e le Septembre 2025
\usepackage[T1]{fontenc}
\usepackage[utf8]{inputenc}
\usepackage[frenchb]{babel}
\usepackage{graphicx}
\makeatletter
\def\@begintheorem#1#2{\trivlist
      \item[\hskip \labelsep{\bf #1\ #2}]\rm}
\def\@opargbegintheorem#1#2#3{\trivlist
      \item[\hskip \labelsep{\bf #1\ #2\ (#3)}]\rm}
\makeatother 
  
\title{A result from the thesis of the french astronomer Alexandre V\'eronnet: critical latitudes (working document, version 1)\\
Sur un r\'esultat de la th\`ese de  l'astronome fran\c cais Alexandre V\'eronnet : les latitudes critiques (document de travail, version 1)}
\author{Herv\'e  Le Ferrand \footnote{h.leferrand@gmail.com, membre de la Soci\'et\'e Math\'ematique de France}}
\date{\today}

\begin{document}

\maketitle

{\bf Abstract}\\
We are interested in a result found in the thesis of the astronomer Alexandre Véronnet (1876-1951) : "Precession of a fluid ring rotating along a parallel axis. Zone of maximum crustal compression at 35 degree. A cause of earthquakes". In 1912, Alexandre Véronnet hypothesizes that the Earth's rotation influences the occurrence of earthquakes in the 35th parallel zone. Has this hypothesis been taken up by other authors, geophysicists, seismologists, etc.? Has it been confirmed or refuted? What is the literature on this subject? Our aim is not to draw a conclusion on this hypothesis but to examine the dissemination of a scientific result.

\section{Latitudes critiques}

En 1912, Alexandre V\'eronnet (1876-1951) soutient,  \`a la Facult\'e des Sciences de Paris (Sorbonne), une  th\`ese intitul\'ee " {\it Rotation de l'ellipso\"ide h\'et\'erog\`ene et figure exacte de la terre} ". Le jury est compos\'e de  l'astronome Henri Andoyer (1862-1929), pr\'esident, d'Henri Poincar\'e (1854-1912), rapporteur,  et de Paul Appell (1855-1930). La th\`ese de V\'eronnet est publi\'ee dans le tome 8 (1912) du {\it Journal de Math\'ematiques Pures et Appliqu\'ees} \cite{Veronnet1912b}, pages 331-463. Elle est consacr\'ee aux {\it figures d'\'equilibre} et notamment \`a la question de l'aplatissement aux pôles. Mais c'est plus pr\'ecis\'ement au paragraphe $5$ du chapitre VI, " Hypoth\`eses autres que celle de Clairaut. Terre solide et vitesse interne variable", de la th\`ese \cite{Veronnet1912a,Veronnet1912b} que nous allons nous int\'eresser. Dans cette partie\footnote{Pages 429 \`a 434 du m\'emoire publi\'e dans le {\it Journal de Math\'ematiques Pures et Appliqu\'ees}.}, intitul\'ee " Pr\'ecession d'un anneau fluide tournant suivant un parall\`ele. Zone de compression maximum de l'\'ecorce \`a $35^{\circ}$. Une cause de tremblements de terre", V\'eronnet \'ecrit \`a la page 429 :
\begin{quotation}
\begin{it}
La d\'eviation due \`a la pr\'ecession\footnote{ " La Terre se comporte comme une toupie en fin de course. Son axe de rotation d\'ecrit un cône autour de la direction perpendiculaire au plan de son orbite (\'ecliptique)."\ (source : https://www.astropolis.fr/espace-culture/dico-astro/astronomie-dico-astro.htm)} et \`a la nutation\footnote{" Oscillation p\'eriodique de faible amplitude se superposant au mouvement de pr\'ecession de l'axe de rotation terrestre"\  (source : https://www.astropolis.fr/espace-culture/dico-astro/astronomie-dico-astro.html)} n'est pas la m\^eme sur tous les parall\`eles d'une m\^eme surface de niveau. Il suffit d'\'etudier cette action sur un anneau fluide formant un parall\`ele quelconque, et consid\'er\'e comme tournant tout d'une pi\`ece en vertu de la rigidit\'e gyrostatique.
\end{it}
\end{quotation}
Plus loin, Alexandre V\'eronnet \'etablit la formule num\'erot\'ee (114') où $\psi$ est l'" arc compt\'e sur l'\'ecliptique"\ et $\frac{\partial \psi}{\partial t}$ est la " vitesse de d\'eplacement autour de l'\'ecliptique"\ :
$$
\left(\frac{\partial \psi}{\partial t}\right)_{P}=h\left(\frac{3}{2}-\frac{1}{\cos^{2}l}\right)\frac{\cos\theta}{\omega}.
$$
puis poursuit :
\begin{quotation}
\begin{it}
On en d\'eduit que la d\'eviation de l'axe de rotation d'un fluide, tournant suivant un parall\`ele, est maximum \`a l'\'equateur, puis toujours d\'ecroissante. Elle s'annule sur le parall\`ele de $35^{\circ}$ où $\cos^{2}l=\frac{2}{3}$ et change de signe ensuite.
\end{it}
\end{quotation}

De ce r\'esultat, Alexandre V\'eronnet \'emet donc l'hypoth\`ese que la rotation de la Terre a une influence sur la survenue des s\'eismes dans la zone du  $35^{\circ}$ parall\`ele. Il \'ecrit en effet :
\begin{quotation}
\begin{it}
Elle [la d\'eviation \`a la surface de la Terre] produirait une acc\'el\'eration tangentielle, dans le sens du m\'eridien, de $4$ cm par seconde, pressant les couches superficielles l'une contre l'autre avec une force \'egale au $\frac{4}{1000}$ de leur poids environ (...) Cette pression lat\'erale va diminuant en latitude suivant la formule (114'), jusqu'au $35^{\circ}$, où elle s'annule et change de signe.

Dans le cours d'une rotation de $24$ heures, les diff\'erents parall\`eles tendront donc \`a se comprimer sur celui de $35^{\circ}$ en deux points diam\'etralement oppos\'es, puis $12$ heures apr\`es \`a s'en \'eloigner. Il y aura compression et dilatation successives tout au long des deux parall\`eles de $35^{\circ}$. Ces pressions lat\'erales convergentes se traduiront aussi par une r\'esultante dirig\'ee suivant le rayon vecteur, comme une onde de mar\'ee diurne. Les parall\`eles de $35^{\circ}$ seront ainsi des zones de fracture et de dislocation de l'\'ecorce (...)

Il suffit de faire remarquer que ce parall\`ele de $35^{\circ}$ passe par San Francisco, le Haut-Mexique, Lisbonne, la Sicile, la Calabre, la Perse, le Japon, qui est bien une ligne privil\'egi\'ee de tremblements de terre. Dans l'h\'emisph\`ere sud, ce parall\`ele se trouve en plein oc\'ean, sauf les pointes de continent : Le Cap, Melbourne, Buenos-Ayres.
\end{it}
\end{quotation}

Alexandre V\'eronnet n'a pas effectu\'e de mesures. Sa remarque d\'ecoule de calculs r\'ealis\'es \`a partir d'une certaine mod\'elisation de la Terre. On peut se demander cependant ce qui a motiv\'e sa r\'eflexion, en cette ann\'ee 1912, sur l'origine des tremblements de terre. Cette hypoth\`ese a-t-elle \'et\'e reprise par d'autres auteurs, g\'eophysiciens, sismologues,... ? A-t-elle \'et\'e confirm\'ee ou infirm\'ee ? Qu'en est-il de la litt\'erature sur ce sujet ? Notre propos n'est pas d'\'emettre une conclusion sur cette hypoth\`ese mais d'examiner la diffusion d'un r\'esultat scientifique. 

Dans un m\'emoire publi\'e en 1914\footnote{Et aussi en 1926, dans \cite{Veronnet1926b}.}, " La forme de la Terre et sa constitution interne"\ \cite{Veronnet1914}, Alexandre V\'eronnet reprend ses analyses :
\begin{quotation}
\begin{it}
(...) on sait qu'un ellipsoïde peu aplati et qui se d\'eforme, comme par exemple sous l'action des mar\'ees de l'\'ecorce, s'articule pr\'ecis\'ement autour des parall\`eles de $35^{\circ}$, qui restent fixes, car ils forment l'intersection avec la sph\`ere de m\^eme volume. Pour cette double raison , et en particulier pour ce jeu alternatif de compression et d\'ecompression, ces parall\`eles deviendront des zones de fracture et de glissement de la mosaïque superficielle, et l'on entrevoit une cause possible d\'eterminante des tremblements de terre.
\end{it}
\end{quotation}

Les r\'esultats de V\'eronnet sur l'aplatissement de la Terre sont cit\'es d\`es 1915 par Willem de Sitter (1872-1934) \cite{Sitter}. W. de Sitter mentionne aussi la latitude $35^{\circ}$\footnote{Les travaux de W. de Sitter sont cit\'es par W. Schweydar dans " Die Bewegung der Drehachse der elastischen Erde im Erkörper und im Raume", {\it Astronomische Nachrichten}, Bans 2023, Nr 4855, 7., 1916.}.

Dans une conf\'erence sur les migrations polaires donn\'ee en 1965 devant la Soci\'et\'e Belge d'Astronomie \cite{Dauvillier}, l'astronome et physicien français Alexandre Dauvillier (1892-1979)\footnote{voir : https://expositions-virtuelles.univ-toulouse.fr/expos/au-laboratoire-d-alexandre-dauvillier/un-homme-de-passions} mentionne ce r\'esultat de la th\`ese d'Alexandre V\'eronnet :
\begin{quotation}
\begin{it}
Le r\'eajustement du sph\'ero\"ide ne s'effectue pas d'une mani\`ere insensible et continue, mais par saccades, chaque fois que les contraintes exc\`edent la limite \'elastique de la lithosph\`ere. Lorsqu'une sph\`ere d\'eformable se transforme, par rotation, en un ellipso\"ide aplati aux pôles et renfl\'e \`a l'\'equateur, elle se d\'eforme de part et d'autre des parall\`eles $35$ degr\'es $16$ minutes, comme l'a fait remarquer A. V\'eronnet en 1912. La sph\`ere et le sph\'ero\"ide de m\^eme volume ont en commun ce parall\`ele invariable.
\end{it}
\end{quotation}

En 2012, Ostrihansky \cite{Ostri} \'ecrit :
\begin{quotation}
\begin{it}
Consideration of the Earth’s rotation as a factor influencing the Earth’s surface is based
on very old data. Already Darwin (1881) recognized that owing to the Earth’s rotation,
the equatorial regions are subjected to greater forces than the polar regions. Böhm von Böhmersheim (1910) presented an opinion that the Earth’s rotation and its changes is
an energy source of orogenetic processes. Next I mention authors Veronnet (1927), Schmidt (1948) and Stovas (1957). The development of Earth’s rotation theories begins
after the confirmation of Earth’s rotation variations by comparison with the atomic
clock and later by exact measurements using very long baseline interferometry (Munk
and Mac-Donald, 1960).
\end{it}
\end{quotation}
Dans cet extrait Ostrihansky mentionne l'ouvrage de V\'eronnet, " Constitution et \'evolution d’univers"\footnote{Gaston Doin et Cie., Paris, pages 62–63, 1927.} paru en 1927, dans lequel est reprise l'hypoth\`ese faite en 1912. Comme le signale Ostrihansky, plusieurs auteurs, \`a partir de la fin des ann\'ees 1940,  s'int\'eressent au lien entre s\'eismes et  variation de la vitesse de rotation de la Terre. Par exemple, Stovas \'ecrit en 1957 \cite{Stovas}:
\begin{quotation}
\begin{it}
To the question of whether the energy of the earth's rotational retardation and of nutation is sufficient for the conjugate deformation of the ellipsoid for the accumulation of stresses leading to a tectonic pulse, and for the formation of folding, an answer is provided by two special researches, one of which is due to L.S. Leibenzon  and the other to the well-known French mathematician A. V\'eronnet.
\end{it}
\end{quotation}
puis :
\begin{quotation}
\begin{it}
In 1912 A.V\'eronnet's doctoral dissertation was published, in which he showed that precession sets up, in the crustal layer,tangential stresses
in the meridional direction, compression and tension toward $\pm 35^{\circ}$. 
\end{it}
\end{quotation}

Stovas donne ensuite l'extrait de la th\`ese de V\'eronnet que nous avons reproduit plus haut. Un peu plus tôt, en 1952, Van den Dungen, Cox et Van Mieghem,  en introduction \`a \cite{VandenDungen},  indiquent :
\begin{quotation}
\begin{it}
Il semble que les tremblements de terre seraient les plus fr\'equents lors de " changements du signe de la marche de
l’horloge Terre".
\end{it}
\end{quotation}

Toujours en 2012, dans une note aux Comptes Rendus de l'Acad\'emie des Sciences de Paris \cite{Rebetsky2012}, Rebetsky dans son panorama des recherches russes en tectonophysique mentionne les \'etudes sur la rotation de la Terre :
\begin{quotation}
\begin{it}
It is crucial to understand the relative impacts
of deviatoric component of gravitation stress tensor, of the
planetary stresses related to the Earth’s rotation (Leibenzon,
1955; Stovas, 1975) (...)
\end{it}
\end{quotation}
et cite deux auteurs que nous allons retrouver plus loin.

Dans un article paru en 2015, " The empirical scheme of short-term prediction of earthquakes and the critical parallels of the Earth", Doda, Malashin, Natyaganov et Stepanov livrent l'analyse suivante \cite{Doda} :
\begin{quotation}
\begin{it}
In the late 1960s the old hypothesis of mobility of the continents by A.Vegener received a
powerful impulse and development, first in the form of kinematic scheme of movement of
lithospheric plates on the underlying mantle. Since then this scheme became a dominant concept of
global tectonics. Many supporters of this concept connect activation of tectonic forces, which
manifests itself in seismic and volcanic activity, with the processes inside the Earth. In this
concept, which many researchers call a theory, in fact there is no place to account rotational forces
at the expense of own rotation of the planet, as well as external influences of the Moon, the Sun and
other planets on the Earth that moves in complex way in space.
At the beginning of XX century the alternative approach was recognized based on the laws of
Newtonian mechanics and taking into account rotational force from the own rotation of the Earth
and the gravitational influence of the Moon and the Sun. This approach allowed to receive
theoretical justification of the so-called critical parallels that were originally partially revealed by
geographical topography of the planet. Significant contribution to the study of critical parallels
made Liebenson (Russia, 1910), A. Veronnet (France, 1912), P. Appell(France, 1932), F.
Krasovsky (USSR, 1941), B. Lichkov (USSR, 1944), M. Stovas (USSR, 1959-1975) and other. The
last author was one of the representatives of the Leningrad school of planetary geophysics
(astrogeology) and he investigated almost the entire set of critical parallels or latitudes of the globe
in the works [8]: $0^{\circ}$ the equator, $\pm 19^{\circ}$, $\pm 35^{\circ}$, $\pm 48^{\circ}$, $\pm 62^{\circ}$, $\pm 90^{\circ}$ – the poles. Some authors distinguish
two critical latitudes ($\pm 45^{\circ}$ and $\pm 65^{\circ}$), and two major critical meridians: the belt of $100^{\circ}$ - $105^{\circ}$ E and
$70^{\circ}$ - $75^{\circ}$ W, $15^{\circ}$ E and $165^{\circ}$ W.
\end{it}
\end{quotation}

Doda et ses collaborateurs, comme Ostrihansky, mettent en exergue les travaux de Stovas.

Plus r\'ecemment, deux auteurs (et leurs collaborateurs) citent les conclusions de V\'eronnet. Levin fait r\'ef\'erence \`a la th\`ese de V\'eronnet dans \cite{Levin2014} et \cite{Levin2017}, et emploie l'expression de " latitude critique". Dans le paragraphe intitul\'e " Analysis of the relationship between the compression
(ellipticity) of a celestial body and its angular rotation rate and other parameters"\ de \cite{Levin2017}, Levin \'ecrit :
\begin{quotation}
\begin{it}
At those indicated latitudes $\pm\varphi_{0}$ ($\varphi_{0}=35^{\circ}15' 52''$), the special feature of which was noted by the French mathematician A. Veronnet[15], variations of the angular rotation rate of a planet are very small, while the surface curvature is independent of the ellipticity and coincides with that for a sphere (Fig. 4).
\end{it}
\end{quotation}
Dans la conclusion de cet article de 2017, Levin affirme :
\begin{quotation}
\begin{it}
In this work, for the first time comparison has been performed
of the observable evidence on the geodetic ellipsoid dynamics
and theoretical estimates obtained from studies of the pulsation
model of the Earth's shape due to variations of its rotation rate.
\end{it}
\end{quotation}

Dans une s\'erie d'articles portant sur la recherche de causes structurelles \`a certains s\'eismes, Eppelbaum et ses collaborateurs \cite{Eppelbaum2017, Eppelbaum2020, Eppelbaum2023, Eppelbaum2024} se r\'ef\`erent \`a la th\`ese de V\'eronnet. Par exemple, dans l'article de 2020, les auteurs \'ecrivent :
\begin{quotation}
\begin{it}
A relationship between the rotation factors, middle latitudes and global geodynamics
is emphasized in [102] [103]. A possible tectonic origin of the discovered
structure (superplume?) may be linked to its critical $\pm 35^{\circ}$ latitude caused by
variations in the Earth’s rotation velocity and tidal forces [104] [105]. Obviously,
these effects producing geoid pulsations can be accompanied by corresponding
changes in the total planet’s volume and, consequently, triggering deformations
and stresses of the Earth. This phenomenon was predicted in the V\'eronnet
theorem [106] indicating that at approximately the $35^{\circ}$ latitude a zone of the
conjugate deformation of the Earth’s ellipsoid could occur. It should be underlined
that the center of the recognized deep structure practically coincides with
the $35^{\circ}$ latitude (Figure 2, Figure 4, Figure 6, Figure 7).
\end{it}
\end{quotation}
La bibliographie comporte plusieurs r\'ef\'erences aux travaux de Levin (et ses collaborateurs).

Dans l'article de 2023, il est indiqu\'e au sujet du r\'esultat de 1912 :
\begin{quotation}
\begin{it}
V\'eronnet’s (1912) comprehensive physical-mathematical analysis of the Earth’s rotating
ellipsoid indicated that the two Earth’s critical latitudes are $\pm 35^{\circ}$. These effects are due to Earth’s
rotation velocity changes and the tidal forces’ uneven impact (V\'eronnet, 1912). Further studies using
extensive geological-geophysical materials demonstrated that, following V\'eronnet’s theory, periodic
matter fluxes in the Earth’s mantle move from the equatorial to the polar regions and vice versa
(Andersson, 2007; Khain and Koronovsky, 2007). Eppelbaum et al. (2021) showed that the latitude of
$+35^{\circ}$ coincides with the center of the projection of the revealed quasi-circular mantle structure (Figure
2).
\end{it}
\end{quotation}

Nous pouvons aussi, comme travaux r\'ecents, mentionner  les \'etudes de Rebetsky et celles de Fodor. En 2016, Rebetsky \cite{Rebetsky2016} note :
\begin{quotation}
\begin{it}
The author continues to investigate additional planetary-level stresses that occur in the crust due to distributed tangential mass forces. Such forces may be related to the daily rotation of the Earth and movements of the relatively solid core relative to the geocenter. In [Rebetskii, 2016], he discusses how the tangential mass forces in the continental crust are influencing additional meridional and latitudinal stresses and attempted to explain regularities of planetary fracturing. In this paper, he considers the role of the tangential mass forces in the occurrence of lateral movement of the lithospheric plates.
The author proposes to estimate amplitudes of the tangential mass forces from the difference between the two global ellipsoids of rotation (...)
The estimates in this study suggest that the tangential mass forces can be viewed as a possible source of the movements of the lithospheric plates.
\end{it}
\end{quotation}

Fodor \cite{Fodor} indique en 2019 :
\begin{quotation}
\begin{it}
In this paper, we analyse the mutual interrelation between earthquake activity and
Earth rotation. The influence of earthquakes on the Earth rotation has been the subject of several
studies before (Varga et al., 2005; Bizouard, 2005; Gross et al., 2006; Xu et al., 2014). Based
on our investigations we concluded that the relationship between these two phenomena could be
detected in the reverse direction too: changes in the speed of Earth’s rotation (that is, changes in
the Length-of-Day (LOD)) may affect earthquake activity.
\end{it}
\end{quotation}

Anatolii Tserklevych, Yevhenii Shylo et Olha Shylo en 2019 \cite{Tserklevych} proposent :
\begin{quotation}
\begin{it}
The purpose of this work is to show how redistribution of masses occurs as a result of gravity-rotational and endogenous forces in the evolutionary self-development of the planet, which leads to the transformation of the lithosphere from the sphere to the biaxial and then to triaxial ellipsoid, and vice versa; and changes in compression and the movement of the pole in geological time. Determine the deformation of the figure of the lithosphere due to the reorientation of the figure's pole.
\end{it}
\end{quotation}
Dans l'introduction, les auteurs faisant un historique des recherches, mentionnent :
\begin{quotation}
\begin{it}
The Earth’s rotation is the most important factor
determining the parameters of the equilibrium shape
of the planet. The peculiarities of the rotation motion
provide information about the internal structure of the
Earth, and the variations of the rotational regime
(speed of rotation and movement of the pole) are a
real source of energy for tectogenesis. The
mathematical reasoning of the role of this factor in
tectogenesis was realized in numerous works by
mathematicians A. Verona [V\'eronnet], P. Appel [Appell], L. S. Leibenzon,
geodesist M. V. Stovas and others
\end{it}
\end{quotation}

En 2020, Belashov \cite{Belashov}, conclut son \'etude sur la d\'eformation conjugu\'ee de l'ellipsoïde lors de la variation de la vitesse angulaire de rotation par :
\begin{quotation}
\begin{it}
From formula (15), which expresses a change of the radius vector with a change of the polar
compression of the ellipsoid, one can see that at $35^{\circ} 21' 18.5''$ $\displaystyle{\frac{\partial r}{\partial\alpha}=0}$, that is, in the zone of this
latitude the radial displacements do not occur with a conjugate change of the figure (...)

In the terminology of Stovas (1957), the zone $\phi=\pm (30-40)^{\circ}$ is the zone of " critical parallels".
\end{it}
\end{quotation}

Nous pouvons aussi citer les travaux de Chen \cite{Chen2010} et ceux de Li (et ses co-auteurs) \cite{Li2021}. Chen conclut son article par :
\begin{quotation}
\begin{it}
The earth basic dynamic system should include the content in two aspects, namely the mantle convection, relating to the earth heat and gravity, and the earth rota-tion. The mantle convection is the power source of the plate break and movement, while the earth rotation re-strains the growing trends, and in turn controls the direction of the plate movement. Above both are two parts of dynamic system that can not be separated.
\end{it}
\end{quotation}

Li \'ecrit :
\begin{quotation}
\begin{it}
Relationship between the occurrence of earthquakes and the Earth’s rate of rotation
has been investigated since the late 1960s. It has been found that there is a significant
correlation between the irregular variation of the rate at which the Earth
rotates and seismic energy released by strong earthquakes of medium and deep
sources(Stoyko and Stoyko 1969) .
\end{it}
\end{quotation}
et affirme dans la conclusion :
\begin{quotation}
\begin{it}
It is no doubt that not only the variation in the Earth’s rotation rate but also the
meteorological processes can cause variational stresses in the crust. The variational
stresses may be too weak to trigger earthquakes.
\end{it}
\end{quotation}

\newpage
\section{El\'ements biographiques sur Alexandre V\'eronnet}

Nous donnons quelques \'el\'ements biographiques sur Alexandre V\'eronnet \cite{LeFerrand}. 

Alexandre V\'eronnet  soutient une th\`ese en Astronomie en 1912. Pr\^etre, Alexandre V\'eronnet quitte l’Eglise \`a la fin de la Premi\`ere Guerre mondiale, conflit durant lequel  premier conflit il est infirmier d’Août 1914 \`a F\'evrier 1919. Cela l’emp\^echera de mener la carri\`ere d’astronome \`a l’Observatoire de Paris qu’il avait d\'ebut\'ee en F\'evrier 1914. 

Alexandre V\'eronnet est l’auteur de nombreuses publications dans les domaines de l’Astronomie, de la Cosmogonie et de la M\'ecanique Rationnelle. V\'eronnet participe en 1930 \`a la r\'edaction du premier fascicule du Tome 4 du Trait\'e de M\'ecanique Rationnelle de Paul Appell (1855-1930). Il en r\'edige enti\`erement en 1937 le second fascicule. Certains r\'esultats de sa th\`ese " Rotation de l’ellipsoïde h\'et\'erog\`ene et figure exacte de la terre"\  soutenue en 1912 \`a la Facult\'e des Sciences de Paris sont toujours cit\'es. Le jury \'etait compos\'e de l’astronome Henri Andoyer (1862-1929), de Henri Poincar\'e (1854-1912), rapporteur, et de Paul Appell. 

Alexandre V\'eronnet est n\'e le 11 Mai 1876 \`a Chagny. Son p\`ere, L\'eonard (1843-1923), est entrepreneur de chars \`a bœufs. V\'eronnet est \'el\`eve au coll\`ege Sainte Marie \`a Chagny, tenu par les Fr\`eres Maristes des Ecoles. Vers 1896, Alexandre V\'eronnet est \'el\`eve eccl\'esiastique \`a Autun. En 1900, Alexandre V\'eronnet alors vicaire \`a Roman\`eche-Thorens, est nomm\'e professeur \`a Semur en Brionnais, puis en Août 1905, il devient professeur au Petit S\'eminaire d’Autun. 

En 1903, Alexandre V\'eronnet est licenci\'e \`es sciences Math\'ematiques et Physique de la Facult\'e des Sciences de l’Universit\'e Catholique de Lyon. Il a notamment comme professeur le comte Magnus de Sparre (1849-1933) auteur de r\'esultats importants dans l’utilisation des fonctions elliptiques \`a la r\'esolution d’\'equations diff\'erentielles. En 1913, Alexandre V\'eronnet b\'en\'eficie d’une bourse de la fondation Commercy pour poursuivre ses recherches en Astronomie et le 4 F\'evrier 1914, il est nomm\'e astronome stagiaire \`a l’Observatoire de Paris. 

Lors de la Premi\`ere Guerre, il rencontre une infirmi\`ere, Jeanne Clerc, qui deviendra son \'epouse en 1919. Il rompt ses vœux. Alexandre V\'eronnet est envoy\'e en mission \`a l’Observatoire de Strasbourg le 1er Mai 1919. Par un arr\^et\'e du 19 Novembre 1919, Alexandre V\'eronnet est de plus nomm\'e charg\'e de conf\'erences de m\'ecanique rationnelle \`a l’Universit\'e de Strasbourg. Il seconde le math\'ematicien Henri Villat (1879-1972). Alexandre V\'eronnet devient astronome-adjoint \`a l’Observatoire de Strasbourg. En 1927, tout en conservant son poste \`a l'Observatoire de Strasbourg, il succ\`ede \`a Henri Villat \`a l’Ecole R\'egionale d’Architecture de Strasbourg pour l’enseignement de la Statique et de la R\'esistance des mat\'eriaux. 

Le {\it Jahrbuch über die Fortschritte der Mathematik}  recense 45 publications d'Alexandre V\'eronnet, articles et ouvrages, entre les ann\'ees 1912 \`a 1935. D’autres articles sont r\'epertori\'es dans {\it Astromiches Jahresbericht}. 

Le 2 Septembre 1939, Alexandre V\'eronnet, son \'epouse et leur fille, doivent quitter Strasbourg pour Clermont-Ferrand. En 1941, il prend sa retraite et s’installe en Bourgogne, \`a Chassey-le-Camp\footnote{Une culture  du N\'eolithique, le Chass\'een, vers 4350-3300 av. J.-C, tire son nom du site de Chassey (https://www.chassey-le-camp.com/histoire-et-prehistoire/le-site-archeologique/presentation-du-site/)}.

\end{document}